\documentclass[a4paper,10pt]{article}
\usepackage{times}%
\usepackage[T1]{fontenc}%
\usepackage[latin1]{inputenc}%
\newcommand{\ie}{{\em i.e.}~}

\newcommand{\Ss}{\mathcal{S}} 
\newcommand{\Ps}{\mathcal{I}} 
\newcommand{\xb}{\mathbf{x}} 
\newcommand{\Ab}{\mathbf{A}} 
\newcommand{\Cb}{{C}}
\newcommand{\SE}{\mathtt{SE}}
\newcommand{\sv}{\mathbf{s}} 

\usepackage[numbers,square]{natbib}
\usepackage{graphicx}
\usepackage[colorlinks=false]{hyperref}
\title{Feature detection using spikes: \\ the greedy approach.%
}%
\author{Laurent Perrinet\\ Institut de Neurosciences Cognitives de la Méditerranée\\(INCM-
UMR 6193, CNRS)\\ 31, ch. Joseph Aiguier, 13402 Marseille Cedex 20, France.\\
\texttt{Laurent.Perrinet@incm.cnrs-mrs.fr}
\\ Tel. : +33-04 91 16 45 23, Fax : +33- 04 91 77 93 04 
}
\date{}
\begin{document}
\maketitle
\begin{abstract}
A goal of low-level neural processes is to build an efficient code extracting the relevant information from the sensory input. It is believed that this is implemented in cortical areas by elementary \emph{inferential} computations dynamically extracting the most likely parameters corresponding to the sensory signal. %
We explore here a neuro-mimetic feed-forward model of the primary visual area (V1) solving this problem in the case where the signal may be described by a robust linear generative model. This model uses an over-complete dictionary of primitives which provides a distributed probabilistic representation of input features. %
Relying on an efficiency criterion, we derive an algorithm as an approximate solution which uses incremental \emph{greedy} inference processes. This algorithm is similar to 'Matching Pursuit' and mimics the parallel architecture of neural computations. We propose here a simple implementation using a network of spiking integrate-and-fire neurons which communicate using lateral interactions. %
Numerical simulations show that this \emph{Sparse Spike Coding} strategy provides an efficient model for representing visual data from a set of natural images. Even though it is simplistic, this transformation of spatial data into a spatio-temporal pattern of binary events provides an accurate description of some complex neural patterns observed in the spiking activity of biological neural networks.
\end{abstract}
\textbf{Keywords: } \textit{Neuronal representation, inverse linear model, over-complete dictionaries, distributed probabilistic representation, spike-event computation, Matching Pursuit, Sparse Spike Coding.}
\section{Toward a functional model of the neural code}
A major problem in neuroscience is to understand the content of the activity that is observed in biological neurons. These complex activity patterns that are the basis of our cognitive abilities remain a mystery and there is yet no known unifying model explaining the "language" that could be used by neurons at the various scales of the central nervous system. In particular, descriptive models of the neural activity tend to be incomplete or to reflect a distorted description of natural conditions \citep{Olshausen04}. We will try here to overcome these problems by precisely defining the model and the hypotheses that we want to validate. We will assume here that there exists a functional \textit{neural code} and that we may decipher the neural activity by exploring algorithms ---based on the nature and architecture of the neural system--- that solve efficiently the function that is provided by the system. We will illustrate this method for the primary visual area (V1) in the human by trying to define precisely its function and then by proposing a model for the neuronal representation and for the mechanisms that may implement it. 
\subsection{Solving inverse problems using neural networks}
\begin{figure}[h]
\centerline{\includegraphics[height=7cm]{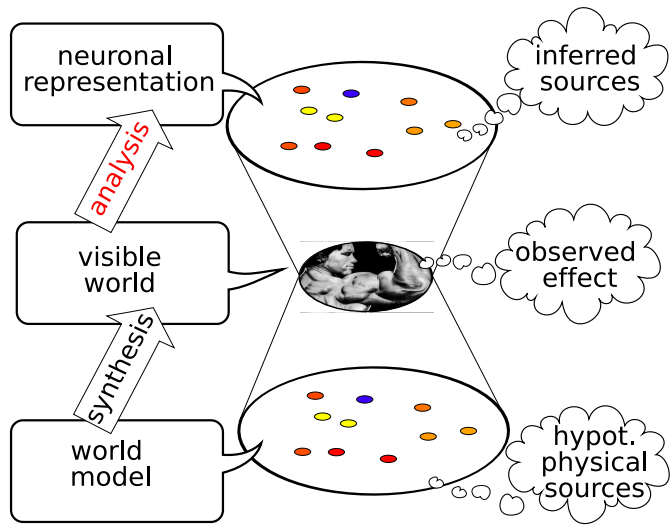}}
\caption{\textbf{Inverse-mapping as a a goal for sensory neural coding}. The visible world is modeled as the interaction of a large set of hypothetical physical sources (world model) according to a known model of their interactions ("synthesis"). We will consider that for sensory cortical areas, the goal of the neural representation (and its implementation by the \emph{neural code}) is to analyze the signal so as to recover at best and as quickly as possible the sources that generated the signal ("analysis") . The analysis may thus be considered as an inverse mapping of the synthesis. A proposed solution for this problem is to \emph{infer} at best the most probable hidden state. }%
\label{fig:inverse}
\end{figure}

V1 is a cortical area specialized in low-level visual processing from which the majority of the visual information diverges to higher visual areas. We will describe it here as implementing an inverse problem by \emph{analyzing} images thanks to an internal model. The hypothesized function over the long term (in the order of hours to years) will thus be to process natural scenes (that is images that occur frequently) so as to progressively build a "model" of their structure. The goal is that for any of these images, this model must rapidly (in the order of a fraction of a second) represent a set of features relevant to that image\footnote{We will consider here that each neuron may be characterized by a preferred pattern to represent. It should though be emphasized that this view differs from the "grand-mother neuron" paradigm since the representation emerges from the interaction of different active neurons.} and corresponding to this model (see Fig.~\ref{fig:inverse}). This representation, including for instance the location and orientation of the edges that outline the shape of an object, is then relayed to higher level areas to allow, for instance, a robust recognition of useful patterns. Actually, this is similar to numerous tasks in engineering and applied mathematics, where a {reverse-engineering} process allows to find a robust representation of the data (such as an estimation of the internal state of a system in control theory) by identifying the so-called \emph{hidden parameters} of the system. The success of this algorithm over the long term  (in the order of days to generations) allows then to validate the model that was learned through the pressure of evolution. In this framework, it is thus easier to describe cortical activity as the result of the inversion (or analysis) of an internal model of the world.\\
Moreover, such a model of the world should also take into account some basic knowledge of actual physical interactions. This idea is based on the assumption that the observations are the effect of the interplay between different causes corresponding to stable physical interactions and that they should be recovered to describe the observed data by representing the underlying actual physical structure. In particular, some  knowledge of the usual transforms of the signal (such as translation and scaling in images) which are related to regularly occurring changes in the physical world (lateral and frontal translations of objects in space) allows then for a robust representation and further analysis by higher level cortical areas. This may finally allow for desired properties such as an invariant representation of objects to these frequently occurring transformations.\\ %
We will restrict here this artificial neural network to a feed-forward model of V1 which processes flashed static images\footnote{In particular, we will study the transient response of the network and neglect the information fed back by higher areas. This latter information will be necessary in more complex algorithms which take into account the context of a local feature.}. We will assume that the model of natural images is fixed and accurate and we  will define the goal of our model as recovering the sources (corresponding to some hidden state variables, see Fig.~\ref{fig:inverse}) from an observed static image. Moreover, in the framework of natural living systems, we will assume that a main constrain from evolution is the ability to process the information as quickly as possible.  This model will consist in these restrictive conditions to a one-layered neural network as illustrated in Fig.~\ref{fig:perceptron} and the output of the neural layer should describe at best and as quickly as possible the visual content. Considering the system as an information channel (according to the definition of \citet{Shannon64}) which processes samples from the set of natural images, we  may therefore define the goal of V1 as to transmit the information about the sources (in \emph{bits}) at the highest rate  as possible.%
\subsection{Inverse models for sensory processing}%
To build an algorithm of the inverse model to efficiently code the input, we will first define the forward synthesis model as a \textit{Linear Generative Model} (LGM) as is often assumed for natural images \citep{Olshausen98}. %
For visual data, images consist indeed of the set of observed luminance values from different spatial positions and a fairly good approximation ---especially for small images of non-occluding objects--- considers the image as the linear combination of "primitive images", similarly to the superposition of transparent layers. This approximation is based on the assumptions that the energy of the photonic flow from a spatial position (the luminosity) consists of the multiplicative interaction of different "shapes" that contribute each for a fraction of the global luminosity. Thanks to the non-linear \emph{gamma} transform of luminosities into luminances \citep{Poynton93} which approaches a logarithmic function, these "shapes" add up linearly in the luminance space. Although this is justified in practice for transparent shapes, it is not for occlusions. The LGM framework provides however a general framework for describing natural images.\\ %
The forward model defines images as the superposition of shapes of different intensities which correspond in our framework to scalar "hidden states". Formally, we will describe this set of scalars by a vector\footnote{in the following, we will denote vectors and matrices by bold characters} $\sv=\{ s_j  \}_{1\leq j \leq N}$  where $N$ is the dimension of the dictionary. Similarly, one image will be described as a point in a multidimensional state space of dimension  $M$  where every pixel corresponds to one dimension (and therefore the pixel value will be its scalar value along this dimension). This observation signal will be written $\xb=\{ x_i \}_{1\leq i \leq M}$ over the set of spatial positions denoted by their address $i$ (that is the pixels over a rectangular grid in an image processing framework). To define the LGM, we will use a "dictionary" of images as the matrix $\Ab=\{ \Ab_j;  1\leq j \leq N\}$ of the $N$ images of the "primitive shapes" $\Ab_j=\{ A_{ij} \}_{1\leq i \leq M}$. The image corresponding to the internal state $\sv$ will finally be defined as:
\begin{equation}
\xb=\sum\nolimits_{1\leq j \leq N} s_j . \Ab_j 
\label{eq:neural_code}%
\end{equation} 
This model of natural images is defined by the statistics of the sources $\Ss$ and by the dictionary $\Ab$ of primitive images. The latter corresponds to the set of basis functions which describe the space  of all observed natural images $\Ps=\{  \xb \}$ that we wish to characterize. \\%
In this paper, we will use the same fixed dictionary of filters (that is $\Ab$) and assume similar hypotheses on the statistics of  $\Ss$  to rate the efficiency of different {coding} strategies. Using this formalization, the function of the neural network consists in recovering the sources by inverting the synthesis process. The results of this inversion (in the space of the neural representation) will thus share the same dimension (that we noted $N$) as the space of the sources, that is the cardinal of the dictionary. As a first approximation (and as is observed in simple cells from V1), the dictionary of primitive shapes will correspond to localized orientation selective edges at different positions and scales resembling Gabor functions \citep{Jones87,Ringach02} at different spatial scales.  This may be particularly adapted in an information theoretic based framework as these shapes correspond to independent features in natural scenes \citep{Bell97independent}. We choose here that the forward model will be described by a wavelet transform \citep{Mallat98} and we will use this architecture to compare different coding strategies.
\subsection{Efficient coding of natural images}
In fact, particular care should be put on the parameters of this wavelet architecture. In particular, it is desirable for  the representations of natural images to be robust to natural conditions. As is the case for natural images, we will consider that the observed signals are generated by sources that share certain features which differ by continuous transformations such as edges at different time, position, orientation or scale. Since the corresponding spatial transformations (translations, rotations and scaling) are very common, if there exists a corresponding transformation in the source space (that is if this transformation of all sources are in the dictionary), the resulting representation of the transformed image should simply be derived by a transformation (in the source space) of the original representation.  Thus, it is necessary for the dictionary to be invariant according to these usual transformations for the representation to be robust. In particular, this allows for instance for higher level areas to detect some specific inputs with an invariance to usual transformations. Typically, this robustness constraint implies in our architecture that the tiling of the wavelet filters is smoother than an orthogonal representation \citep{Perrinet03ieee}. As a consequence, the dictionary will be over-complete, \ie the number of dictionary elements will be of several orders of magnitude larger than the dimension of the image space (that is $N >> M$). \\%
From the definition of the forward model, for any signal $\xb$, there exist at least one set of parameters $\sv$ which recovers the observed signal. However, in the case where the dictionary is over-complete, the inversion of the LGM will not yield an unique solution in $\Ss$ to any given signal in  $\Ps$ : the problem is ill-posed. The coding strategies corresponding to possible 'analysis' algorithms (see Fig.~\ref{fig:inverse}) have different efficiencies and, in particular, the solution given by the wavelet coefficients with an over-complete dictionary yields an highly redundant representation. According to \citet{Barlow01}, the goal of sensory processing would be rather to choose the most efficient representation: following the same argument as the Occam razor, whenever there is the choice between two representations, the best is the one that is the most parsimonious. In our framework, a possible goal would be to maximize the mean codeword length, that is get the coding strategy that describes at best the images. From Shannon's source coding theorem \citep{Shannon64}, this length is bounded by the entropy of the images for a given architecture and coding strategy. Under some assumptions that we will develop later, this is equivalent to find the \textit{sparsest representation}, that is the representation that uses the smallest number of sources \citep{Olshausen98}. This sparseness constraint thus allows to restrict the different solutions of the inversion of the forward model so as to find an appropriate candidate for the neural code.\\%
However, the combinatorial  complexity of the inverse problem grows very quickly as the dimension of the dictionary increases (it's {NP-complete}, see \citep{Mallat98}). There exists therefore no simple algorithm that optimizes exactly the problem in reasonable time as we handle more complex signals such as natural images, but acceptable sub-optimal strategies to approach this problem do exist (see a review in \citep{Pece01}). Most popular solutions  optimize a compromise between the reconstruction error and the sparsity and are based on linear optimization or gradient approaches \citep{Simoncelli01}. Following the same arguments as \citet{Barlow01}, we explore an  alternate solution which uses a probabilistic representation and Bayesian inference.%
\section{Sparse spike coding using a greedy inference pursuit}
Focusing on the event-based nature of axonal information transduction and in order to reflect the parallel architecture of the nervous system, we will here propose a solution for inverting the forward model that we defined for natural images. This will build a Bayesian inference framework based on feature-matching neurons and on spikes as events representing primitive "decisions".%
\subsection{Greedy inference pursuit using spikes}
This approach proposes an alternative to classical paradigms of neural coding such as the spike-rate coding approach of the \textit{perceptron} (see Fig.~\ref{fig:perceptron}). Instead of coding information in the mean firing frequency of neurons, we will present an original approach solving the function that we defined above. It uses a distributed probabilistic representation and we will assume here that the activity of neurons (such as the membrane potential) in the layer will represent dynamically the evidence of a correct match and that the output spiking signal signifies a set of elementary decisions made by the neurons. Following this process and focusing on every single spike, the process occurs repeatedly using two steps: Matching (M) and Pursuit (P). 
\begin{enumerate}
\item[(\emph{M})] To each neuron is assigned a vector (or weight pattern) corresponding to its preferred stimulus. Neurons compete in parallel to find the most probable \textit{single source} component by integrating evidence according to their weight patterns. The first source to be detected should be the one corresponding to the highest activity. %
\item[(\emph{P})] The best match is assigned a decision which, once it has been taken, is assumed to be reliable: we can take into account this information {before} performing any further computations (and in particular finding a new match) so as to yield a new representation where we removed \textit{completely} the detected source.
\end{enumerate} 
We call this approach  a \textit{greedy pursuit} which is based on the recursion of two greedy mechanisms (detection - removal). These are here idealized but correspond to known aspects of neural activity (matching - suppression).\\
We will see that this method is similar to the approach developed in the method of Matching Pursuit  \citep{Mallat93}. However, instead of a heuristic scheme, the algorithm will here be derived from known hypotheses and thanks to the description of the successive steps that may lead to the greedy pursuit, it may be considered as an optimization strategy of the goal that we defined above (namely maximizing the transfer of information). We will then propose an implementation using Integrate-and-fire neurons and test the efficiency of this artificial neural code.
\subsubsection{Matching: Detection of the most probable source component}
First, given the signal $\xb \in \Ps$, we are searching for the  \textit{single} source $ s^* . \Ab_{j^*} \in \Ps$  that corresponds to the maximum \textit{a posteriori} (MAP) realization for $\xb$ (and knowing it is a realization of the LGM as it is defined in Eq.~\ref{eq:neural_code}). We will address in general a single source by its index and strength by $ \{ j, s \}$ so that the corresponding vector in $\Ss$ corresponds to a vector of zero values except for the value $s$ at index $j$. The MAP is defined by:%
\begin{equation}
 \{ j^*, s^* \} = \mbox{ArgMax}_{\{ j, s\}} P( \{ j, s\} | \xb )
\label{eq:map}
\end{equation}%
 To evaluate  $P( \{ j, s\} | \xb )$, the probability \textit{a posteriori} of a single source knowing the signal, we have from Bayes' theorem %
\begin{equation}
\{ j^*, s^* \} = \mbox{ArgMax}_{\{ j, s\}} [P( \xb | \{ j, s\} ) . P(  \{ j, s\} )]%
\end{equation}
where $P( \xb | \{ j, s\} )$ is the likelihood probability of a signal knowing a single source and $ P(  \{ j, s\} )$  is the  \textit{a priori} probability of the sources.\\%
To compute the likelihood we have to first define the model of the measurement \citep[p.26]{MacKay03}. We will first assume that we are in a low-noise limit environment (the global contrast is optimal and the eye/camera is adapted to the scene) so that we have no or little measurement noise. Knowing one component $\{ j, s\}$, the only "noise" from the viewpoint of neuron $j$ is the combination of the unknown sources $\{ \alpha_k  \}_{1\leq j \leq N}$. It is thus the residual of the signal knowing $ \{ j, s \}$. We may thus write the noise as %
\begin{equation}
 \xb= s . \Ab_{j} +\mathbf{\nu} \mbox{ with }  \nu =\sum\nolimits_k \alpha_k . \Ab_k %
\end{equation}
The residual of the signal (an image) is thus considered as an undetermined perturbation\footnote{It should be stressed that the image model is still deterministic.}. Assuming that the $\alpha_k$ are independent random variables (since we know only $\{ j, s\}$), from the central limit theorem it comes that for a sufficiently high number of sources, the distribution of the random variable $\nu$ converges to a normal distribution with known mean and covariance matrix. From the work of \citet{Atick92}, we know that for natural images this normal distribution is fairly homogeneous across natural images. We may either use another metric (based on the Mahalanobis distance, as exposed in \citep{Perrinet03ieee}) or use a decorrelating kernel to yield a spherical  probability distribution centered around the origin ($ E(\nu) \ = 0$) of this "noise". Normalizing by the mean energy of images in $\Ps$, the residual signal is thus considered as a decorrelated noise of unit variance. From $P( \xb | \{ j, s\} )=P(\xb - s . \Ab_{j} )= P(\nu)$, it follows
\begin{eqnarray}
 \{ j^*, s^* \} &=& \mbox{ArgMax}_{\{ j, s\}} [\log P( \xb | \{ j, s\} ) + \log  P(  \{ j, s\} )]  \nonumber\\
&=& \mbox{ArgMin}_{\{ j, s\}} [\| \xb - s . \Ab_{j} \|^2 /2- \log P(  \{ j, s\} )]
\end{eqnarray}
We will further consider that the dictionary was learned so that over a long period the neurons have similar statistics: the prior is uniform across sources and values. We thus have no prior knowledge or preference for any source. %
It thus comes 
\begin{eqnarray*}
\{j^* ,  s^* \} &=&\mbox{ArgMin}_{\{ j, s\}}  \| \xb - s . \Ab_{j} \|^2  \\
&=& \mbox{ArgMin}_{\{ j, s\}} [ s^2 . \|  \Ab_j \|^2 -2 . s.  <\xb, \Ab_j> ]
\end{eqnarray*}
To minimize this bi-variate function, we may first minimize for every element $j$  the coefficient $s_j$ to get the corresponding $s_j^* = \mbox{ArgMax}_s P( \{ j, s\} | \xb )$. From the above equations, this is equivalent to minimizing in the last equation the quadratic function of $s$ which is minimal for the scalar coefficient
\begin{equation}
s_j^* =\frac{<\xb, \Ab_j>}{\|  \Ab_j \|^2} %
\label{eq:proj}%
\end{equation}
that is for the scalar projection of the input on $\Ab_j$. 
Then,  since for every element $j$,  $  s_j^* . \Ab_{j}$ is  the projection of $\xb$ on $\Ab_j$, so that $ s_j^* . \Ab_j$ and $ \xb - s_j^* . \Ab_{j}$ are orthogonal, it follows from Pythagoras's theorem
\begin{eqnarray}
 j^*  &=& \mbox{ArgMin}_j  [ \| \xb - s_j^* . \Ab_{j} \|^2  ]  \nonumber\\
&=& \mbox{ArgMin}_j  [ \| \xb  \|^2  -   \| \frac{<\xb, \Ab_j>}{\|  \Ab_j \|^2} . \Ab_j \|^2 ]
= \mbox{ArgMax}_j  \| \frac{  <\xb, \Ab_j>  }{\|  \Ab_j \|} \|^2  \nonumber\\
 j^* &=& \mbox{ArgMax}_j   | <\xb, \frac{ \Ab_j }{\|  \Ab_j \|}> | 
\end{eqnarray}
Finally, as defined in Eq.~\ref{eq:map}, we found that the source component that maximizes the probability  is the projection of the signal on the normalized elements of the dictionary. This justifies the computation of the correlation in the perceptron model \citep{Rosenblatt60} as it provides a measure of the log-probability under the assumptions that we used. However, using a different strategy as these linear systems, we will associate in our greedy approach this inference with a lateral propagation of this information to the correlated neurons and only then resume the algorithm. 
\subsubsection{Pursuit: Lateral interaction and Greedy pursuit of the best components}
Before detecting another single source component, we will take into account the information that we extracted from the signal by propagating it to the neighboring neurons using lateral interaction links. As we found the MAP source knowing the signal $\xb$, we may pursue the algorithm by accounting for this inference on the signal knowing the element that we found. From
\begin{equation}
P( \{ j, s\} | \xb , \{ j^*, s^*\} )= P( \{ j, s\} | \xb - {s^*}. \Ab_{j^*} ) %
\end{equation}
and since source are here supposed to have independent activities\footnote{For any realization of the images, individual sources have independent activities, that is that removing one source, one gets a new image (conform with the LGM model) and one does not change the probability distribution of the other sources.}, the pursuit algorithm assumes that ---knowing the previous detection--- we may resume the detection on this residual signal. We will thus  use this {new residual signal} in which we will then find a new component corresponding to the most probable single source.\\
In this recursive approach, we will note as $n$ the rank of the step in the pursuit (which begins at $n=0$ for the initialization). Writing $N_j=\| A_j \|$, the first scalar  projection that we have to maximize ---and which will serve as the initialization of the algorithm---is given by :
\begin{eqnarray}
   C_j^{(0)}=   <\xb, \frac{ \Ab_j }{N_j }>  
\end{eqnarray} 
Let's also note the address of the successive winning neuron from the first step $n=1$ as 
\begin{equation}
j^{(n)}= \mbox{ArgMax}_j   | \Cb_j^{(n-1)}  |%
\label{eq:argmax}%
\end{equation}
Knowing $j^{(n)}$, in order to resume the pursuit at the next step, we saw that we need to compute the projection of the signal on the elements of the dictionary. Let's therefore set initially $\xb^{(0)}=\xb$ and  $\xb^{(n)}$ the successive residuals. In  this greedy approach, we consider that the decision corresponding to the MAP criteria at step $n$ is correct and that we may therefore update the residual  and the corresponding activities $\Cb_j^{(n-1)}  =  <\xb^{(n-1)}, \frac{ \Ab_j }{N_j }> $ by subtracting to $\xb^{(n-1)}$ its projection on the winning element of index $j^{(n)}$ (see Eq.~\ref{eq:proj}) : 
\begin{equation}
\xb^{(n)}=\xb^{(n-1)}  -\Cb_{j^{(n)}}^{(n-1)} .  \frac{\Ab_{j^{(n)}}}{N_{j^{(n)}}} %
\label{eq:rec}%
\end{equation}
Furthermore, we don't need to feed this information back to the signal and we may directly compute the activity again for all vectors  thanks to the linearity of the scalar product operator: 
\begin{eqnarray}
\Cb_j^{(n)} &=&   <\xb^{(n)}, \frac{ \Ab_j }{N_j }> \nonumber\\
&=&  <\xb^{(n-1)} -  \Cb_{j^{(n)}}^{(n-1)}.\frac{\Ab_{j^{(n)}}}{N_{j^{(n)}}}, \frac{ \Ab_j }{N_j }>  \nonumber\\
\Cb_j^{(n)} &=& \Cb_j^{(n-1)} -   \Cb_{j^{(n)}}^{(n-1)} . <\frac{\Ab_j }{N_j },  \frac{ \Ab_{j^{(n)}}}{N_{j^{(n)}}} >
\label{eq:update}%
\end{eqnarray}
In this simplified framework, the choice of the best match and the update rule are independent of the choice of the norm $N_j$ of the filters (see  Eq.~\ref{eq:argmax} and \ref{eq:update}), so that we may indifferently use in the following normalized filters (that is $N_j=1$ for all neurons) so as to simplify the equations. It comes thus:
\begin{eqnarray}
 \mbox{(Initialization)}&\fbox{$
   \begin{array}{rcl}
   C_j^{(0)}=   <\xb,  \Ab_j >  
  \end{array}
$} 
\end{eqnarray} 
This activities' update (Eq.~\ref{eq:update}) corresponds in neuro-physiological terminology to a lateral interaction. It will be proportional  to ${R_{j,{j^{(n)}}} }$ where $R_{j,{j^{(n)}}}=<\Ab_j , \Ab_{j^{(n)}}>$ is the correlation of any element $j$ with the winning element ${j^{(n)}}$ and relates to the reproducing kernel in wavelet theory. \\
Finally, we achieve the recursive greedy pursuit of best components  as the iteration of respectively a "matching" and a "pursuit" step. While the residual energy is greater than a fixed threshold $ \|  \xb^{(n)}\| > \varepsilon$, we compute :
\begin{eqnarray}
 \mbox{(Matching)}&\fbox{$
   \begin{array}{rcl}
j^{(n)}  &=&  \mbox{ArgMax}_j   | \Cb_j^{(n-1)}  | \\
  \end{array}
$} \label{eq:mp1}
\\
\mbox{(Pursuit)}&\fbox{$
 \Cb_j^{(n)} =\Cb_j^{(n-1)} -    \Cb_{j^{(n)}}^{(n-1)}  . R_{j,{j^{(n)}}} 
 $} 
\label{eq:mp2}
\end{eqnarray}
The greedy pursuit therefore transforms an incoming signal $\xb$ in a list of ranked sources $\{ j^{(n)},s^{(n)} \}$ such that finally (from Eq.~\ref{eq:rec}) the signal may be reconstructed as 
$$ \xb = \sum\nolimits_{k=1\ldots n} s^{(k)} . \Ab_{j^{(k)}}  + \xb^{(n)} $$
which is an approximation of the goal set in inverting Eq.~\ref{eq:neural_code} if the norm of the residual signal $ \xb^{(n)} $ converges to zero. 
\subsubsection{Properties of the greedy pursuit}
This algorithm is exactly equivalent to Matching Pursuit \citep{Mallat93}. This algorithm is familiar in signal processing and is increasingly used for image and video processing \citep{Durka04,Capobianco03}. However, the use of the statistics of natural images statistically optimizes the coding efficiency by modifying the image space metric \citep{Perrinet03ieee}. Moreover, the Bayesian inference framework allows to precisely tune the heuristic approach of the Matching Pursuit. It allows for instance to set a different prior or  to include  knowledge of the measurement noise that is adapted to the goal of the system (and hence a different matching criteria that may depend on the $N_j$). This algorithm presents similar computational complexity and properties \citep[pp.412--9]{Mallat98}. In particular 
 \begin{eqnarray}
\Cb_{j^{(n)}}^{(n)}  = \Cb_{j^{(n)}}^{(n-1)} - \Cb_{j^{(n)}}^{(n-1)} =0
\end{eqnarray}
 and as a consequence the activity of a winning neuron is totally canceled.\\
 Moreover, although filters in the dictionary are here generally not orthogonal, the residual image is orthogonal to the winning filter and
\begin{equation}
 \| \xb^{(n)} \|^2 = \| \xb^{(n-1)} \|^2  -|s^{(n)}|^2 .  \| \Ab_{j^{(n)}}  \| ^2 %
\end{equation}
so that we may easily compute the Squared Error (SE) of the residual signal at every step of the coding.
 \begin{eqnarray}
\SE^{(n)}  &=&   \| \xb -  \sum\nolimits_{k=1\ldots n} s^{(k)} . \Ab_{j^{(k)}} \|^2 = \| \xb^{(n)} \|^2 \nonumber\\
                  &=&  \SE^{(n-1)} -  |s^{(n)}|^2 .  \|  \Ab_{j^{(n)}} \|^2 \\
 \SE^{(n)} &=&  \| \xb  \|^2 -  \sum\nolimits_{k=1\ldots n} |s^{(k)}|^2 .  \|  \Ab_{j^{(k)}} \|^2\nonumber\\
 		&=&  \| \xb  \|^2 -  \sum\nolimits_{k=1\ldots n}    |  C^{(k-1)}_{j^{(k)}} |^2 
\label{eq:se}
\end{eqnarray}
It first implies that the stopping criteria may be computed using this computation without computing $\| \xb^{(n)} \|$. A further consequence of the monotonous decrease of the SE from Eq.~\ref{eq:se} is ---under the condition that the dictionary is at least complete--- the convergence of the reconstruction \citep[p.414]{Mallat98}. Under this condition, the algorithm will therefore stop in finite time.\\%
Though simple, the greedy pursuit is a complex non-linear algorithm. In fact, the study of its behavior is non trivial and may involve chaotic dynamics \citep{Davis94}. In particular, it is obvious that the choice that is made at a giving step may influence all future steps. This implies that a failed match may propagate wrong information to following steps and therefore that the probability of a failure grows higher as the rank increases. These properties are discussed in \citep{Perrinet03ieee} and in particular we illustrated that the speed of convergence increases as the dictionary becomes more  over-complete so that it provides an efficient representation for natural scenes in image processing tasks. 
\subsection{Implementation using Integrate-and-Fire (IF) neurons}
\begin{figure}[ht]
\centerline{\includegraphics[width=6cm]{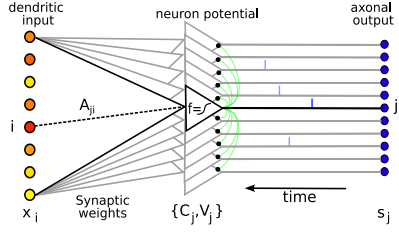}}%
\caption{\textbf{Model of a neuronal layer as a communication channel}. To understand the content of neural activity, we consider here that the neuronal layer implements the inverse of a forward model (that is the analysis in Fig.~\ref{fig:inverse}). The architecture is similar to the perceptron: the input (noted $x_i$) is matched with normalized weight patterns $A_{ji}$ (which are fixed in this paper) so as to provide an integrative activation value (the membrane potential) which in turn is non-linearly transformed to achieve a membrane potential which grows proportionally to the probability of matching a feature. Spikes represent decisions that are fed back on the correlated neighboring neurons using lateral interactions (that we represented for the first spiking neuron) but also on the axonal output which yield a spiking output  $s_j$. } %
\label{fig:perceptron}
\end{figure}

From our knowledge of neural mechanisms in a neuronal layer, the model of greedy feature pursuit that we derived from an event-based computation in a parallel architecture is particularly adapted to a model of neural computations. We will derive an implementation using a network of spiking neurons based on the same  feed-forward architecture of the perceptron (see Fig.~\ref{fig:perceptron}) but implementing the greedy pursuit using \textit{lateral interactions}. \\%
The activity is represented by a driving current that drives the potential $V_j$ of Integrate-and-Fire neurons \citep{Lapicque07}. For illustration purposes, the dynamics of the neurons will here be modeled  by a simple linear integration of the driving current $\Cb_j$ (other integration schemes lead to similar formulations):
\begin{equation}
 \tau . \frac{d}{dt} V_j = p_j.\Cb_j %
\end{equation}
The neurons are duplicated with opposite polarity $p_j=\pm 1$ so that  $\Cb_j=p_j.|\Cb_j|$ to model the ON / OFF symmetry of simple cells \citep{Ringach02}. The neuron will generate a spike  when the potential reaches an arbitrary threshold that we set here to $1$.\\
To implement the computation of the match of an input with stored patterns, we define a dictionary  which will be implemented by weight vectors $\Ab_j$. These vectors are normalized as described above and the input is decorrelated. The linear feed-forward perceptron integrates synaptically the input $\xb$ into an initial activity $C_j$ such that
\begin{equation}
 C_j = <\xb , \Ab_{j}>%
\end{equation}
The scalar projection will therefore drive the potential of the neuron. %
We may  predict from the monotonous integration that the first neuron to generate a spike will be the one that corresponds to the maximal rectified scalar projection of the input signal with the weight vectors of the network, that is
\begin{equation}
j^{*}=  \mbox{ArgMax}_j   | \Cb_j  | %
\end{equation}
the firing time is $t^*=\frac{\tau}{|C_{j^*}|}$ and the potential is then $V_j=\frac{t^*}{\tau}.C_j=\frac{C_j}{|C_{j^*}|}$.
This is therefore a simple and biologically plausible implementation of a MAP estimate using the parallel architecture of the network which is in contrast with the complexity of this implementation on a single-processor computer. To implement the greedy algorithm, we then need to implement a lateral interaction on the neighboring neuron similar to the observed lateral propagation of information in V1 \citep{Grinvald94,Bair03}.  In our scheme the interaction should yield the same configuration in the network (activity and potential) as if the source that was detected was originally absent from the signal. In this model, if $j^*$ is the winning neuron, the activity should be subtracted by $|C_{j^*}|.R_{\{j,j^*\}}$ (see Eq.~\ref{eq:mp2}) and the potential by this value integrated over $t^*$. The lateral interaction is  thus achieved by updating after each spike the activity of the neighboring neurons proportionally to their cross-correlation $ R_{\{j,j^*\}}$ with the corresponding  winning neuron $j^*$ :
\begin{equation}
C_j \leftarrow \Cb_j - |\Cb_{j^*}| .R_{\{j,j^*\}}%
\label{eq:lat1}%
\end{equation}
and removing the potential that would be generated by the activity of the removed source:
$$V_j \leftarrow V_j -   \frac{t^*}{\tau}. |C_{j^*}| .R_{\{j,j^*\}}$$
that is simply
\begin{equation}
V_j \leftarrow V_j -   R_{\{j,j^*\}} %
\label{eq:lat2}%
\end{equation}
This lateral interaction is here immediate and behaves as a refractory period on the winning neuron ($C_{j^*} \leftarrow 0$ and $V_{j^*} \leftarrow 0$) and a graded inhibition on positively correlated neurons. It involves a subtractive hyper-polarizing term on the potential and on the activity. Biologically, it is improbable that the lateral interaction could be instantaneous, but this lateral interaction could be implemented in a fast manner using a shunting lateral interaction \citep{BorgGraham98} mediated by fast-spike inter-neurons. Finally, this simple implementation therefore implements the Matching Pursuit algorithm that we defined in Eq.~\ref{eq:mp1} and \ref{eq:mp2}  and we will apply it to simple visual tasks. 
\section{Results: efficiency of Sparse Spike Coding}
\subsection{Coding natural image patches}%
\begin{figure}%
 \centerline{ \includegraphics[width=5cm]{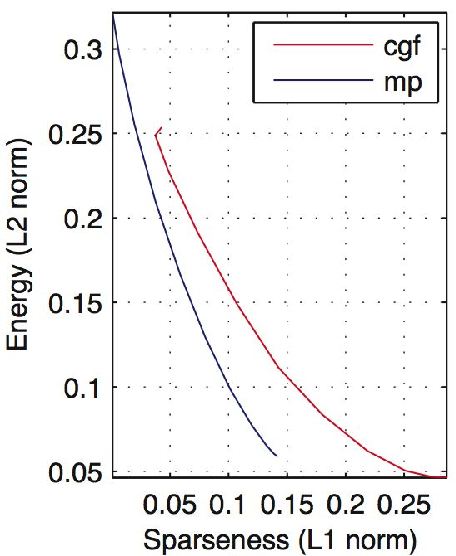}%
 \includegraphics[width=5cm]{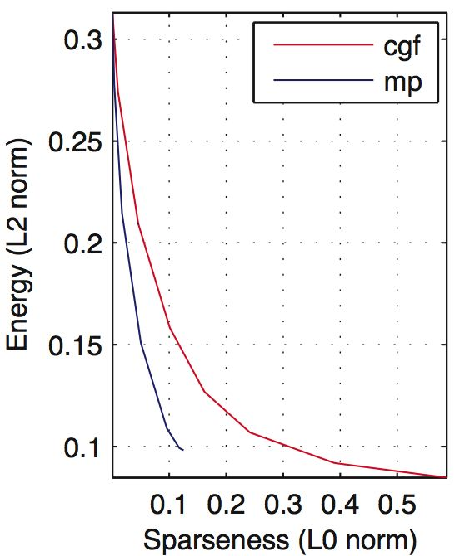}%
}%
 \caption{\textbf{Efficiency of the matching pursuit compared to conjugate gradient}. We compared here the matching pursuit ('mp') method with the classical conjugate gradient function ('cgf') method as is used in \citep{Olshausen98}. We present the results for the coding of a set of image patches drawn from a database of natural images. These results were obtained with the same fixed dictionary of edges for both methods. We plot the mean final residual error for two definitions of sparseness : {\bf (Left)} the mean absolute sum of the coefficients and {\bf (Right)} the number of active (or non-zero) coefficients (the coding step for MP). For this architecture, the sparse spike coding scheme appears to be more efficient to code natural image patches.}
\label{fig:efficiency}
\end{figure}
We compared the method we described in this paper with similar techniques used to yield sparse and efficient codes such as the conjugate gradient method used by  \citet{Olshausen96nature}. We used a similar context and architecture as these experiments  and used in particular the database of inputs and the dictionary of filters learned in the {\sc Sparsenet} algorithm. Namely, we used a set of $10^5$  $10\times10$ patches (so that $M=100$) from whitened images drawn from a database of natural images.  The weight matrix was computed using the  {\sc Sparsenet} algorithm with a 2-fold over-completeness ($N=200$) that show similar structure as the receptive of simple cells in V1 \citep{Ringach02}. From the relation between the likelihood of having recovered the signal and the squared error in the new metric, the mean squared reconstruction error (L2-norm) is an appropriate measure of the coding efficiency for these whitened images. This measure represents the mean accuracy (in terms of the logarithm of a probability) between the data and the representation. We compared here this measure for different definitions and values for the "sparseness". \\
First, by changing an internal parameter tuning the compromise between reconstruction error and sparsity (namely the estimated variance of the noise for the conjugate gradient method and the stopping criteria in the pursuit), one could yield different mean residual error with different mean absolute value of the coefficients (see Fig.~\ref{fig:efficiency}, left) or L1-norm. In a second experiment, we compared the efficiency of the greedy pursuit while varying the number of active coefficients  (the L0-norm), that is the rank of the  pursuit. To compare this method with the conjugate gradient, a first pass of the latter method was assigning for a fixed number of active coefficients the best neurons while a second pass optimized the coefficients for this set of "active" vectors (see Fig.~\ref{fig:efficiency}, right).  \\
Computationally, the complexity of the algorithms and the time required by both methods was similar. However, the pursuit is by construction more adapted to provide a progressive and dynamical result while the conjugate gradient method had to be recomputed for every set of parameter. Best results are those giving a lower error for a given sparsity or a lower sparseness (better compression) for the same error. In both cases, the Sparse Spike Coding  provides a coding paradigm which is of better efficiency as the conjugate gradient.
\subsection{Model of a hyper-column in the primary visual area}
\begin{figure}%
 \centerline{\includegraphics[height=6cm]{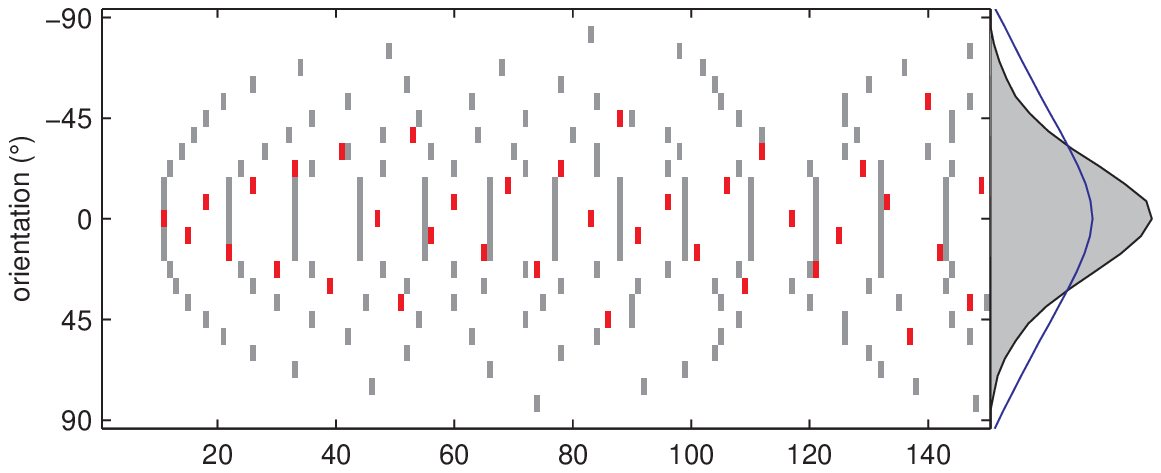}}
 \caption{\textbf{Implementation of the greedy pursuit using Integrate-and-Fire Neurons}. We simulated here the activity of a network of Integrate-and-Fire neurons tuned to form a simple model of an hyper-column in the primary visual area (V1) to the presentation of a horizontal edge at $t=0$. We show in this image the output spiking activity of $16$ neurons tuned for different orientations for the feed-forward (black bars) and the sparse spike coding (white bars) models during the first $150$ ms. In this latter model, the correlation linked to the information already detected is propagated as a hyper-polarizing and shunting lateral interaction to the neighboring neurons : the response in both latency and spiking frequency to the oriented edge is clearly more selective.}
\label{fig:mp_bio}
\end{figure}
\begin{figure}%
 \centerline{\includegraphics[height=6cm]{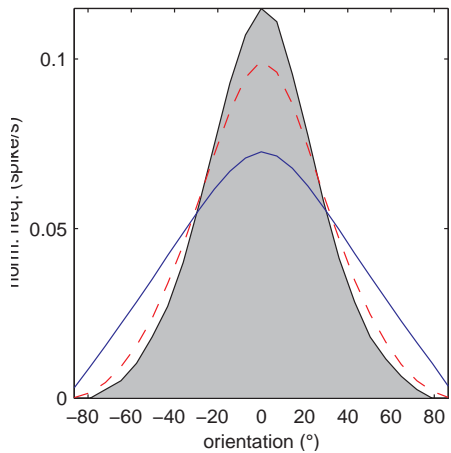}}
 \caption{\textbf{Selectivity response of the network to orientation}. Output spike firing rate to the presentation of a horizontal edge at time $t=0$. for the linear feed-forward model (plain line), the sparse spike coding scheme (filled curve) and with divisive normalization (dashed line) for different orientations of the input stimulus. The narrower tuning curve for the latter two methods represents a more selective response to the features learned in synaptic weights and mimics the behavior of the neural response in the primary visual area.}
\label{fig:mp_bio_prob}
\end{figure}

To illustrate the properties of the algorithm, I modeled a network of linear Integrate-and-Fire neurons forming a simple model of an hyper-column in the primary visual area (V1). This model consist of an isolated network of $16$ neurons selective to different orientations of contours and which are modeled as Gabor filters (which are here symmetric with circular envelopes). We compared a pure feed-forward model to a network implementing the lateral interactions that we described above (see Eq.~\ref{eq:lat1} and \ref{eq:lat2}). We show here the resulting spiking activity when one of the preferred stimuli (the horizontal edge) was continuously presented from time $t=0$ (see Fig.~\ref{fig:mp_bio}).\\
We observe that the neuron corresponding to that preferred stimulus fires with the shortest latency but also produces the highest spike rate. Moreover, the activity of the neurons corresponding to non-preferred directions shows a lower spiking activity when implementing the greedy pursuit. This dynamic reflects the lateral interaction (here an inhibition to the positively correlated neurons) generated at every spike which is observed in V1 \citep{Celebrini93}. In fact, compared to the linear model, the latency  and the frequency of the neighboring neurons show a sharper response for neighboring edge orientations (see Fig.~\ref{fig:mp_bio_prob})  which corresponds to the  high selectivity observed in simple cells from V1 \citep{Ringach02b}. The selectivity of this model was compared with the model of \emph{divisive normalization} \citep{Schwartz01}, suggesting that this simple implementation of Integrate-and-Fire neurons ---linked by lateral interactions and removing dynamically the redundancy in the signal--- could provide a model for the complex processing occurring in cortical areas.
\section*{Conclusion}
We presented here a model for neural processing which provides an alternative to the feed-forward and spike-rate coding approaches.  Focusing on the parallel architecture of cortical areas, we  based our computations on spiking events. Defining the function of sensory areas as matching the input to a model with unknown parameters, the activity of the network represented a probabilistic evaluation of the accuracy of the match. From this representation, we inferred the best match using the Bayes rule and an inference decision criterion. We then derived an algorithm which may be implemented using lateral interactions : it removes for every spike the corresponding activity to correlated neurons. Simulations of this model compare to the non-linear behavior of neurons in biological network such as the primary visual cortex (V1).\\%
This model is based on the Matching Pursuit algorithm and provides a general framework for modeling the complex behavior of networks of spiking neurons. Particularly, it can be extended to multi-layered networks and provides an efficient code for natural images as we described elsewhere \citep{Perrinet03ieee}. Further studies provided a  learning scheme based on an Hebbian learning rule which yields an unsupervised learning of the sources as independent components of the signal to describe \citep{Perrinet04nc}. The model thus provides an algorithm of \textit{Sparse Spike Coding} which is particularly efficient for visual tasks.\\%
This simple strategy thus suggest that the inherent complexity of the neural activity is perhaps not simply the reflection of the computational details of neurons but may rather be the consequence of the parallel event-based dynamics of the neural activity.  Although our model is a simplistic caricature compared to the behavior of biological neurons, it provides a simple algorithm which is compatible with some complex characteristic of the response of neuronal populations.  It thus proposes a challenge for discovering the mechanisms underlying the efficiency of nervous systems by focusing on large-scale networks of spiking neurons.%
\subsubsection*{Reproducible research}
Scripts reproducing all figures may be obtained from the author upon request.
\subsubsection*{Acknowledgments}
The author thanks the team at the Redwood Neuroscience Institute for stimulating discussions and particularly Jeff Hawkins, Bruno Olshausen, Fritz Sommer, Tony Bell, Dileep George, Kilian Koepsell and Matthias Bethge. This work was supported by a grant form the French Research Council  (Action Concertée Incitative / Temps et Cerveau).


\begin{thebibliography}{28}
\expandafter\ifx\csname natexlab\endcsname\relax\def\natexlab#1{#1}\fi
\expandafter\ifx\csname url\endcsname\relax
  \def\url#1{{\tt #1}}\fi

\bibitem[Atick(1992)]{Atick92}
J.~Atick.
\newblock Could information theory provide an ecological theory of sensory
  processing?
\newblock {\em Neural Computation}, 3\penalty0 (2):\penalty0 213--52, 1992.

\bibitem[Bair et~al.(2003)Bair, Cavanaugh, and Movshon]{Bair03}
W.~Bair, J.~Cavanaugh, and J.~Movshon.
\newblock Time course and time--distance relationships for surround suppression
  in macaque {V}1 neurons.
\newblock {\em The Journal of Neuroscience}, 23\penalty0 (20):\penalty0 7690
  ---701, August 2003.

\bibitem[Barlow(2001)]{Barlow01}
H.~Barlow.
\newblock Redundancy reduction revisited.
\newblock {\em Network: Computations in Neural Systems}, 12:\penalty0 241---25,
  2001.

\bibitem[Bell and Sejnowski(1997)]{Bell97independent}
A.~Bell and T.~Sejnowski.
\newblock The `independent components' of natural scenes are edge filters.
\newblock {\em Vision Research}, 37\penalty0 (23):\penalty0 3327--38, 1997.

\bibitem[Borg-Graham et~al.(1998)Borg-Graham, Monier, and
  Fregnac]{BorgGraham98}
L.J.~Borg-Graham, C.~Monier, and Y.~Fregnac.
\newblock Visual input evokes transient and strong shunting inhibition in
  visual cortical neurons.
\newblock {\em Nature}, 6683\penalty0 (393):\penalty0 369--73, 1998.

\bibitem[Capobianco(2003)]{Capobianco03}
E.~Capobianco.
\newblock Independent multiresolution component analysis and matching pursuit.
\newblock {\em Comput. Stat. Data Anal.}, 42\penalty0 (3):\penalty0 385--402,
  2003.
\newblock ISSN 0167-9473.

\bibitem[Celebrini et~al.(1993)Celebrini, Thorpe, Trotter, and
  Imbert]{Celebrini93}
S.~Celebrini, S.~Thorpe, Y.~Trotter, and M.~Imbert.
\newblock Dynamics of orientation coding in area {V1} of the awake primate.
\newblock {\em Vis Neurosci}, 5\penalty0 (10):\penalty0 811--25, 1993.

\bibitem[Davis(1994)]{Davis94}
G.~Davis.
\newblock {\em Adaptive Nonlinear Approximations.}
\newblock PhD thesis, New York University, 1994.

\bibitem[Durka et~al.(2001)Durka, Ircha, and Blinowska]{Durka04}
P.~Durka, D.~Ircha, and K.~J. Blinowska.
\newblock Stochastic time-frequency dictionaries for matching pursuit.
\newblock {\em IEEE Tran Signal Process}, 49\penalty0 (3):\penalty0 507--510,
  March 2001.

\bibitem[Grinvald et~al.(1994)Grinvald, Lieke, Frostig, and
  Hildesheim]{Grinvald94}
A.~Grinvald, E.~Lieke, R.~Frostig, and R.~Hildesheim.
\newblock Cortical point-spread function and long-range lateral interactions
  revealed by real-time optical imaging of macaque monkey primary visual
  cortex.
\newblock {\em The Journal of Neuroscience}, 14\penalty0 (5):\penalty0
  2545--68, May 1994.

\bibitem[Jones and Palmer(1987)]{Jones87}
J.~P. Jones and L.~A. Palmer.
\newblock An evaluation of the two-dimensional gabor filter model of simple
  receptive fields in cat striate cortex.
\newblock {\em J. Neurophysiol.}, 58\penalty0 (6):\penalty0 1233--58, 1987.

\bibitem[Lapicque(1907)]{Lapicque07}
L.~Lapicque.
\newblock Recherches quantitatives sur l'excitation {\'e}lectrique des nerfs
  trait{\'e}e comme une polarisation.
\newblock {\em J. Physiol. (Paris)}, 9:\penalty0 620--35, 1907.

\bibitem[MacKay(2003)]{MacKay03}
D.~ MacKay.
\newblock {\em Information Theory, Inference and Learning Algorithms}.
\newblock Cambridge University Press, 2003.

\bibitem[Mallat(1998)]{Mallat98}
S.~Mallat.
\newblock {\em A wavelet tour of signal processing.}
\newblock Academic Press, 1998.

\bibitem[Mallat and Zhang(1993)]{Mallat93}
S.~Mallat and Z.~Zhang.
\newblock Matching pursuit with time-frequency dictionaries.
\newblock {\em IEEE Transactions on Signal Processing}, 41\penalty0
  (12):\penalty0 3397--3414, 1993.

\bibitem[Olshausen and Field(1996)]{Olshausen96nature}
B.~Olshausen and D.~Field.
\newblock Learning a sparse code for natural images produces a multiscale
  family of localized receptive fields.
\newblock {\em Nature}, 1996.

\bibitem[Olshausen and Field(1998)]{Olshausen98}
B.~Olshausen and D.~Field.
\newblock Sparse coding with an overcomplete basis set: A strategy employed by
  {V1}?
\newblock {\em Vision Research}, 37:\penalty0 3311--25, 1998.

\bibitem[Olshausen(2004)]{Olshausen04}
B.~Olshausen.
\newblock What is the other 85{\%} of {V}1 doing?
\newblock In L.~van~Hemmen T.J.~Sejnowski, editor, {\em Problems in Systems
  Neuroscience}. Oxford University Press, 2004.

\bibitem[Pece(2001)]{Pece01}
A.~Pece.
\newblock The problem of sparse spike coding.
\newblock Technical Report DIKU-TR-2001-02, Institue of computer science,
  University of Copenhagen, Copenhagen, Denmark, 2001.

\bibitem[Perrinet(2004)]{Perrinet04nc}
L.~Perrinet.
\newblock Finding independent components using spikes : a natural result of
  hebbian learning in a sparse spike coding scheme.
\newblock {\em Natural Computing}, 3\penalty0 (2):\penalty0 159--75, January
  2004.
\newblock URL \url{http://laurent.perrinet.free.fr/publi/perrinet03nc.pdf}.

\bibitem[Perrinet et~al.(2004)Perrinet, Samuelides, and Thorpe]{Perrinet03ieee}
L.~Perrinet, M.~Samuelides, and S.~Thorpe.
\newblock Coding static natural images using spiking event times : do neurons
  cooperate?
\newblock {\em IEEE Transactions on Neural Networks, Special Issue on 'Temporal
  Coding for Neural Information Processing'}, 15\penalty0 (5):\penalty0 1164--
  1175, September 2004.
\newblock ISSN 1045-9227.
\newblock URL \url{http://laurent.perrinet.free.fr/publi/perrinet03ieee.pdf}.

\bibitem[Poynton(1993)]{Poynton93}
C.~Poynton.
\newblock Gamma and its disguises.
\newblock {\em Journal of the Society of Motion Picture and Television
  Engineers}, 102\penalty0 (12):\penalty0 1099--108, December 1993.

\bibitem[Ringach(2002)]{Ringach02}
D.~Ringach.
\newblock Spatial structure and symmetry of simple-cell receptive fields in
  macaque primary visual cortex.
\newblock {\em J. Neurophysiology}, 88:\penalty0 455--63, 2002.

\bibitem[Ringach et~al.(2002)Ringach, Shapley, and Hawken]{Ringach02b}
D.~Ringach, R.~Shapley, and M.~Hawken.
\newblock Orientation selectivity in macaque {V}1: diversity and laminar
  dependence.
\newblock {\em J. Neurosci.}, 22\penalty0 (13):\penalty0 5639--51, 2002.

\bibitem[Rosenblatt(1960)]{Rosenblatt60}
F.~Rosenblatt.
\newblock Perceptron simulation experiments.
\newblock {\em Proceedings of the I. R. E.}, 20:\penalty0 167--192, 1960.

\bibitem[Schwartz and Simoncelli(2001)]{Schwartz01}
O.~Schwartz and E.~Simoncelli.
\newblock Natural signal statistics and sensory gain control.
\newblock {\em Nature Neuroscience}, 4\penalty0 (8):\penalty0 819--25, 2001.

\bibitem[Shannon and Weaver(1964)]{Shannon64}
C.~Shannon and W.~Weaver.
\newblock {\em The mathematical theory of communication.}
\newblock The University of Illinois Press, Urbana, 1964.

\bibitem[Simoncelli and Olshausen(2001)]{Simoncelli01}
E.~Simoncelli and B.~Olshausen.
\newblock Natural image statistics and neural representation.
\newblock {\em Annual Review of Neuroscience}, 24:\penalty0 1193--216, 2001.

\end{thebibliography}
\end{document}